# Ultrafast modelocked nonlinear micro-cavity laser

M. Peccianti[1,2,*], A. Pasquazi[1], Y. Park[1], B. E. Little[3], S. T. Chu[3,+], D. J. Moss[1,4,**] and R. Morandotti[1]

[1]*INRS-EMT, 1650 Blvd. Lionel Boulet, Varennes (Québec), J3X 1S2 Canada.*

[2]*IPCF-CNR, UOS Roma and ISC-CNR UOS Montelibretti, Via dei Taurini 19, 00185 Roma, Italy.*

[3] *Infinera Ltd, 169 Java Drive, Sunnyvale, California 94089, USA.*

[4]*CUDOS, School of Physics, University of Sydney, Sydney, NSW, Australia 2006.*

[+]*present address: Department of Physics and Materials Science, City University of Hong Kong*

[**] *D.J.Moss present address: School of Electrical and Computer Engineering, RMIT University, Melbourne, Victoria, Australia 3001*

*\* alessia.pasquazi@gmail.com*

**Ultra-short pulsed lasers, operating through the phenomenon of mode-locking, have played a significant role in many facets of our society for 50 years – for example in the way we exchange information, measure and diagnose diseases, process materials and in many other applications. The ability to phase-lock the modes of the high-quality resonators recently exploited to demonstrate optical combs, would allow mode-locked lasers to benefit from their high optical spectral quality in order to realize novel sources such as precision optical clocks for applications to metrology, telecommunication, microchip-computing, and many other areas. We demonstrate the first mode-locked laser based on a micro-cavity resonator. It operates via a new mode-locking method we termed Filter-Driven (FD) Four-Wave-Mixing, and is based on a CMOS-compatible high quality factor micro-ring resonator. It achieves stable self-starting oscillation with negligible amplitude noise at ultrahigh repetition rates, and spectral linewidths well below 130 kHz.**

Passively mode-locked lasers have generated the shortest optical pulses to date [1]. Self-supporting ultra-short optical pulses can be produced naturally in these lasers [1-4] by the complex nonlinear interactions between chromatic dispersion, the Kerr nonlinearity and saturable gain [5,6]. There is considerable interest in achieving both very high and flexible repetition rates, at frequencies well beyond the range of active mode-locking where the use of electronics poses intrinsic limitations. Many different approaches have been proposed to achieve this, ranging from very short laser cavities with large mode frequency spacings (i.e., large *Free*

*Spectral Range* (FSR) [1,7-8], where a very high repetition rate is achieved by simply reducing the pulse round-trip time (at the expense of line quality), to schemes where multiple pulses are produced in each round trip [5,9-10], applied primarily to fibre lasers. A pioneering approach to the latter was introduced in 1997 by Yoshida et al. [10], where a Fabry Pérot (FP) filter inserted in the main cavity suppresses all but a few modes that are periodically spaced, leading to a train of pulses with a controlled repetition rate. This approach was subsequently reinterpreted in terms of the dissipative Four-Wave-Mixing (FWM) paradigm [11,12], and since then, high repetition-rate pulse trains have been demonstrated by adopting variations of this method [12-15]. Although in principle dissipative FWM schemes allow for transform limited pulses to be generated at repetition rates not limited by the main laser cavity, their stability still remains a severe issue - a common problem [1] when multiple pulses circulate in a cavity. This is a consequence of the fact that the gain and nonlinearity required to sustain proper lasing action necessitate the use of very long fibre cavity lengths. This in turn produces smaller cavity mode frequency spacings of (typically) a few MHz or less, which allows many modes to oscillate within the FP filter bandwidth. Since these modes have random phases, their beating results in severe low frequency noise that produces extremely unstable operation [14].

In this paper, we report the first mode-locked laser based on a nonlinear monolithic high-Q (quality factor) resonator. Our laser achieves extremely stable operation at high repetition rates while maintaining very narrow linewidths, thus leading to a high quality pulsed emission. The key point is that this resonator is not simply used as a filter but acts as the nonlinear element as well [16], similar in spirit to optically pumped multiple wavelength oscillators based on high Q-factor resonators [16-22]. Due to this twofold central role of the nonlinear filter, we term this new mode-locking scheme *Filter-Driven Four-Wave-Mixing* (FD-FWM). It operates in a way which is in stark contrast to traditional dissipative FWM schemes where the nonlinear interaction occurs in the fibre and is then "filtered" separately by a *linear* FP filter. Our method has two immediate advantages, specifically it is (i) *intrinsically more efficient;* in the alternate approach the nonlinear mixing is usually performed right after the amplification stage, in the point of the loop where the pulse energy is maximum. The linear filtering follows the nonlinear step, cutting part of the spectral broadening. In our design the filtering cannot be considered to be separated from the nonlinear process. As some frequency components simply do not survive in the filter due to destructive interference, they are not able to seed the energy transfer process toward forbidden bands in the spectrum, consequently reducing the general power loss. It (ii) *drastically reduces the main cavity length, which substantially increases the main cavity mode frequency spacing*, thus dramatically increasing the laser stability, as (ideally) only one cavity mode is now allowed to oscillate within the ring resonance. The combination of these factors enables our scheme to achieve stable operation at high repetition rates over a large range of operating conditions, *without any additional stabilization system.* In addition, our FD-FWM scheme can intrinsically produce much narrower linewidths than ultrashort cavity mode-locked lasers (that oscillate at similar repetition rates) because the long main cavity results in a much smaller Schawlow-Towns phase noise limit [23,24].

Yet a further key advantage of our approach lies in the fact that the stable lasing regime is highly robust to external (i.e., thermal) perturbations, that are a well-known problem in resonator-based optical parametric oscillators (OPOs) [17,22]. In OPOs, temperature variations due to the circulating power naturally detune filter resonances with respect to an external pumping source. Although stabilization mechanisms such as *thermal locking* have been exploited to solve this problem, they are ineffective against slow temperature drifts, a fact which often leads to the situation where the OPO shuts down [25]. Our approach does not suffer from this problem as all the oscillating lines are actually modes of the main cavity that lase within the resonator resonances. As the nonlinear resonator is part of main cavity, the changes in the resonator optical path induced by heating also contribute to a change in the main cavity length, which moderates the frequency misalignment between ring and main cavity modes. Moreover the stable oscillation is sustained primarily by external gain and it is obtained at resonator input power levels in the 10mW range, about an order of magnitude lower than the values normally used to pump OPOs. The ring resonator heating issues are therefore reduced. However, we found that the central lasing frequency was dependent on the energy coupled in the ring, which yielded a variation of the central frequency over the power at the input of the ring of approximately 0.67pm/mW.

The key component of the laser, *the ring resonator*, shown in Figures 1 (a-c), is an integrated micro-ring resonator with a Q-factor of 1.2 million [21,26] (estimated from a measured linewidth of 160MHz), fabricated in a CMOS-compatible chip platform [27,28] based on a high-index, doped-silica glass waveguide. The experimental setup is shown in Figure 1 (b)

(see the Methods section for details). The micro-ring resonator is simply embedded in a standard Erbium doped-fibre loop cavity (acting as the gain medium) containing a passband filter with a bandwidth large enough to pass all of the oscillating lines, with the main purpose of controlling the central wavelength $\lambda_0$. A delay line is employed to control the phase of the main cavity modes with respect to the ring modes.

In order to investigate the dependence of the laser performance on the main cavity length we conducted experiments with two lasers, one based on a short-length Erbium-Doped fibre amplifier (EDFA) and the other based on a long high-power Erbium-Ytterbium fibre amplifier (EYDFA), the latter having a similar operating regime to dissipative FWM lasers, where long fibre based cavities are necessary to assure the Kerr nonlinearity required for modelocked laser operation. The two configurations therefore had significantly different main cavity lengths (3 and 33m), i.e. different FSRs (68.5MHz and 6MHz respectively) as well as different saturation powers.

Figure 2 shows the experimental optical spectra of the pulsed output (Figure 2 (a,c)) along with the temporal traces obtained by a second order noncollinear autocorrelator (Figure 2 (b,d)) for the two systems (EYDFA on the left and EDFA on the right) at four input powers to the ring resonator. The pulses visible in the autocorrelation trains have a temporal duration that decreases noticeably as the input power increases, as expected for a typical passive mode-locking scheme.

From these plots it would appear that the laser based on the high-power EYDFA had superior overall performance since its pulsewidth was considerably shorter. However, this ignores the fact that the autocorrelation measurements average over any long time scale fluctuations of the laser, without requiring a stable pulsed output. The key issue of laser stability or line coherence is better illustrated by a comparison between the experimental autocorrelation traces with the calculated traces (Figure 2 (b,d) green plots) for a fully stable and coherent system possessing the optical spectra in Figure 2 (a,c). While a perfect match is found for the short length EDFA case, the long cavity design shows a considerably higher background, thus clearly distinguishing unstable from stable laser operation [22].

By changing the amplifier driving current we observed a minimum lasing threshold of 500 and 650mA for the unstable and stable designs respectively (Figure 2 (e-f)) that corresponded to a gain of approximately 20dB for the both cases. For the stable case, the input power in the ring vs the driving current is plotted in Figure 2(f). To preserve the stable operation the delay line was adapted at the mode-locking threshold. This yielded a slight change in total losses that induced a kink in the plot around a current of 850mA. The maximum input power in the ring was 15.4mW for the stable case, limited by the performance of the EDFA.

To better quantify the pulse-to-pulse amplitude stability of the two lasers we recorded the electrical radio-frequency (RF) spectrum of the envelope signal, collected at the output using a fast photo-detector. Unstable oscillation (in the pulse amplitude) was *always* observed for the long cavity design for the EYDFA (Figure 3 (a)) for both CW and pulsed regimes, due to the presence of a large number of cavity modes oscillating in the ring resonance. In complete contrast to this, the short-cavity configuration for the EDFA (Figure 3 (b,c)) could easily be stabilized to give the very clean result of Figure 3 (c), by simply adjusting the main cavity length in order to centre an (ideally) single cavity mode with respect to the ring resonance, thereby completely eliminating any main cavity low-frequency beating. Several stable oscillation conditions were found through tuning the delay by over 2 cm. Note that for the same gain, the optical bandwidth (small black insets in Figure 3 (b,c)) for unstable laser operation was wider (i.e. leading to shorter output pulses) because the instability resulted in a strong amplitude modulation of the optical pulse train in the main cavity, thus increasing the statistical peak power and enhancing the nonlinear interactions. This is a common occurrence in mode locked lasers [29].

Although we expect that significant external temperature changes would eventually detune the external cavity length, once the thermal stability was reached, standard environmental conditioning (21°C) was sufficient to maintain perfectly stable operation within the time-framework of our experimental sessions (several hours), highlighting the robustness of the stability condition.

We confirmed, by numerical simulations (see the Methods Section for details), the dependence of the stability on the relative phase between the main cavity modes and the ring resonator modes (Figure 4). We found that instability could indeed be either induced or suppressed simply by optimizing the phase of the main cavity modes, i.e. the spectral position of the main cavity modes relative to the ring resonances.

We concluded our experiments by characterizing the phase noise [7,24,30-32]. We found that all of the emitted lines had a bandwidth well below 130 kHz (FWHM). This bandwidth has been found to be dramatically increased by the contribution of the environmental acoustic noise that we could minimize but not fully eliminate. Following the approach of Ref. [24], we discriminated the linewidth associated with the inherent bandwidth-quantum-limit, estimated to be below 13kHz, which suggests that, ultimately, proper packaging and acoustic isolation would reduce the optical spectral linewidth even further. We also investigated the source timing jitter, i.e. the stochastic deviation of the temporal pulse position in the train. Following Refs. [7, 30-32] we estimate from the phase noise that the jitter contribution to the RF spectral lines is < 10 KHz . This is a remarkable result for a 200GHz mode-locked source that does not use any auxiliary stabilization mechanisms.

Our ultimate objective is a versatile, fully monolithically integrated laser source. Since waveguide amplifiers have been demonstrated in silica glass platforms, we do not envisage any fundamental issues preventing the full integration of our scheme. This proof-of-concept device represents a key step in realising a fully integrated, stable, high–performance, laser source operating at flexible and very high repetition rates.

In summary, we propose and demonstrate a novel mode-locked laser based on a monolithic high-quality (high-Q) resonator, capable of generating both picosecond and sub-picosecond transform-limited pulses at a repetition rate of 200.8 GHz and beyond. Our device operates via a new mechanism that enables *stable* mode-locked lasing with negligible amplitude noise, and extrinsically limited phase noise, not constrained by the repetition rate. We believe this work represents a key milestone in the generation of ultra-stable, high repetition rate, optical pulse sources, particularly because of its CMOS-compatible monolithic platform.

**Methods**

**Simulations.** The numerical simulations were performed by solving the coupled equations of the field evolution $f(z,t)$ in the amplifying fibre (denoted by subscript $F$) and of the field $a(z,t)$ evolution in the nonlinear ring (denoted by subscript $R$)

$$\partial_z f(z,t) + \frac{n_F}{c}\partial_t f(z,t) + i\frac{\beta_{2F}}{2}\partial_{tt} f(z,t) = g(f)\left(1 - \frac{\partial_{tt}}{\Omega}\right) f(z,t) - \alpha f(z,t)$$

$$\partial_z a(z,t) + \frac{n_F}{c}\partial_t a(z,t) + i\frac{\beta_{2R}}{2}\partial_{tt} a(z,t) + i\gamma |a(z,t)|^2 a(z,t) = 0$$

(1)

Here $c$ is the speed of light, $n_F$, $n_R$ and $\beta_{2F}$, $\beta_{2R}$ are the group indices and second order dispersion in the fibre ($F$) and in the ring ($R$), respectively. Also $\gamma$ is the Kerr nonlinear coefficient in the ring, while for the fibre ($F$), $\alpha$ is the linear absorption and $g(f)$ is the saturable gain expressed as

$$g(f) = G_0 \left[ Exp\left(-\int_0^L |f(z,t)|^2 \frac{dz}{P_0 L}\right)\right]$$

(2)

with $G_0$ and $P_0$ representing the fibre low-signal gain and the amplifier saturation power -which controls the energy in the laser cavity -, and $L$ is the main cavity length. We also note that $\Omega$ regulates the fibre gain bandwidth. The equations are solved in time with a pseudo-spectral method and are coupled at the ring ports according to the relation

$$\begin{pmatrix} a_{out} \\ f_{out} \end{pmatrix} = \begin{pmatrix} t & r \\ -r & t \end{pmatrix} \begin{pmatrix} a_{in} e^{i\Phi_{RC}} \\ f_{in} e^{i\Phi_{MC}} \end{pmatrix}$$

(3)

$t$ and $r$ are the transmission and reflection coefficients that regulate the ring bandwidth, $\Phi_{MC}$ dictates the position of the main cavity modes with respect to the ring modes, and $\Phi_{RC}$ fixes the position of the ring modes with respect to the centre of the gain bandwidth.

The simulations were performed for a 200GHz repetition rate system, starting from noise and letting the system reach the stationary state. The main cavity is shorter when compared to the experiment, in order to fit the problem within the 32Gbyte memory of our multiprocessor system. The operation is investigated for a ring line $FWHM_{RC}$ =30GHz and a main cavity $FSR_{MC}$ =12GHz. The ratio $FWHM_{RC}/FSR_{MC}$ = 2.5 has been chosen to match the experimental conditions. The dispersion is scaled accordingly to obtain a total dispersion corresponding to the experiments.

In Figure 4 (a-f), the simulation results (output spectrum and amplitude) are plotted for growing amplifier saturation energy (from blue to black), for $\Phi_{MC}=0$ (Figure 4 (a-c)) and $\Phi_{MC}=0.875\pi$ (Figure 4 (d-f)). The cavity energy affects the stability as it regulates the number of main cavity modes that oscillate (Figures 4 (b,e)). Configurations with lower energy and modes centred exactly on a ring resonator resonance ($\Phi_{MC}=0$ rad) are typically stable. The parameter used in the experiments to distinguish the stable and unstable regimes is the bandwidth of the RF spectrum. In the numerics it corresponds to the low frequency components (<100GHz in the simulations) of the spectrum of the temporal power waveform $|f_{out}(t)|^2$. In Figure 4 (g) the FWHM of the RF spectrum versus $\Phi_{MC}$ is shown for different amplifier saturation powers. This quantity is expressed in the units of main cavity modes (operatively the FWHM is divided by the FSR of the main cavity), i.e. $FWHM_{RF}=1$ $MC_{modes}$ represents a bandwidth as large as the FSR, and it corresponds to a moderate unstable behaviour, e.g. the red line in Figure 4 (a-c). As it is visible, in Figure 4(g) the stable operating range in terms of phase delay $\Phi_{MC}$ shrinks with the cavity energy.

**Device.** The waveguides used to fabricate the high-Q ring resonators in our experiment possess a high effective nonlinearity (220 $W^{-1}$ $km^{-1}$) due to a combination of tight mode confinement and high intrinsic Kerr coefficient ($n_2$ is 5 times that of silica glass). The dispersion is optimized for FWM in the C-band as it is small and anomalous (~-10$ps^2$/m) near $\lambda$= 1550nm [26], guaranteeing a large FWM gain when pumped in the C-Band. Most significantly, the waveguides exhibit very low linear (<0.04 dB/cm) and negligible nonlinear optical loss (i.e., two photon absorption) up to 25GW/$cm^2$. All of these factors make this platform particularly attractive for low power nonlinear optics in the C-band [21, 26-28]. The input and output bus waveguides of the resonator are pigtailed to standard SMF fibre using integrated mode converters and a V-groove technology, resulting in coupling losses of < 1.5dB/facet.

**Set up and experiments.** The micro-ring resonator is embedded in a larger fibre loop cavity, containing an optical amplifier to provide gain, a Faraday isolator, a polarization controller and a free space delay line. A passband filter, with a bandwidth large enough to pass all of the oscillating lines, is used in order to control the central wavelength $\lambda_0$. The optical amplifiers used to perform the experiments reported here were either an short-length erbium-doped amplifier (EDFA from PriTel) or a more conventional (longer length) high-power Erbium-Ytterbium doped fibre amplifier (EYDFA). The EYDFA had a maximum saturation output power of ~3W, and produced a main cavity FSR of 6.0MHz, corresponding to a main cavity fibre length of 33m, while the short-length EDFA had a saturation output power of ~35mW and resulted in a cavity length below 3m, corresponding to a very large main cavity FSR of 68.5MHz. For the EYDFA based laser, the bandpass filter in the fibre loop used to control the central oscillating wavelength was a 15nm wide filter with ~ 4dB loss. The filter imposes a gain difference of approximately 0.5dB between the central and the first side modes. For the short-length laser a 5nm wide filter was used instead, characterized by a loss of 1.5dB. We verified that the use of different filters for each laser had no impact on the performances in terms of overall stability or linewidth - the filter bandwidth was chosen to optimize the performance of each laser. For the EYDFA based laser, much higher pump powers were available and so decreasing the loss was not as important as providing a wide enough bandwidth to achieve the shortest possible pulsewidth. For the short laser, a limited gain was available and hence the low-loss, narrow bandwidth filter was chosen – also because the overall output bandwidth was much smaller than the EYDFA laser due to the lower available cavity energy. The central lasing wavelengths of the two lasers were adjusted in order to obtain the maximum achievable bandwidth. The different optimal operating wavelength is mainly related to the different main cavity dispersion (anomalous in both cases) and different amplifier gain profiles.

**Amplitude Noise Characterization.** The laser amplitude noise was determined by measuring the electrical radio-frequency (RF) spectrum of the laser using a fast photodetector connected to an amplifier (bandwidth ~ 200MHz).

For the stable case (Figure 3(c)) the RF signal exhibited a dominant DC component with a bandwidth < 0.25Hz (the resolution of our measurement system) as well as an out-of-band noise 55dB lower than the DC peak. We estimate the ratio between the power of the DC component and spectral noise (within the 200MHZ bandwidth) to be > 41dB, again limited by the sensitivity of the measurement, which in our case is dictated by detection and sampling noise. For the unstable case (Figure 3 (a)) the RF bandwidth is very large and the DC component brings actually a quite weak spectral contribution, the ratio always being << 0dB. We estimate this to be ~ -13dB for the 6MHz FSR main cavity.

**Phase Noise Characterization.** To characterize the phase noise, we measured the linewidth of the oscillating lines by first selecting each of them with a narrowband filter and then by analysing the output using a re-circulating delayed self-heterodyne interferometer (RDSHI) with a delay line of 5.8km, and a modulation frequency of 6.5MHz [7,24,30-32]. As reported in Figure 5, we found that all of the observable emitted lines had a bandwidth well below 130 kHz (FWHM) – only limited by the environmental acoustic noise, a factor that we could minimize but not completely eliminate. This is consistent with our best-fit estimation of the intrinsic Lorentzian contribution to the linewidth being < 13kHz, i.e. the quantum limit for the linewidth [24], and implies that ultimately, proper packaging and acoustic isolation would significantly reduce the RF spectral linewidths. Notice that the widths of the emission lines of the unstable design were exactly matching the ring bandwidth as also confirmed by the output amplitude spectrum.

Although it is not possible to obtain the pulse jitter amplitude directly from the RF spectrum [31] due to the extremely high repetition rate, under the assumption that the phase and time jitter noises are white and uncorrelated [32] the spectral broadening $\Delta w_{jitter}$ of the RF lines due to the timing jitter is known to be a parabolic coefficient of the optical spectral broadening $\Delta w_n$ of the laser lines with respect to the central line: $\Delta w_n = \Delta w_0 + \Delta w_{jitter}(N - N_0)^2$ where $N$ is a generic mode number, $N_0$ is the mode number corresponding to the minimum linewidth, and $\Delta w_0$ accounts for other phase noise contributions [7]. As the resolution of our RDSHI measurements was <10kHz, we estimate a pulse jitter bandwidth $\Delta w_{jitter}$ <10kHz.

**Acknowledgements:** This work was supported by the Canadian FQRNT and NSERC, and the Australian Research Council (ARC) Centres of Excellence and Discovery Projects programs. M. P. acknowledges the support of the Marie Curie People project TOBIAS PIOF-GA-2008-221262. We also acknowledge support from QPS Photonics, Enablance and MPB Communications. Finally, we would like to thank Stefano Trillo for the discussions on the theoretical basis of nonlinear dissipative systems and Tudor Johnston for a critical revision of the manuscript.

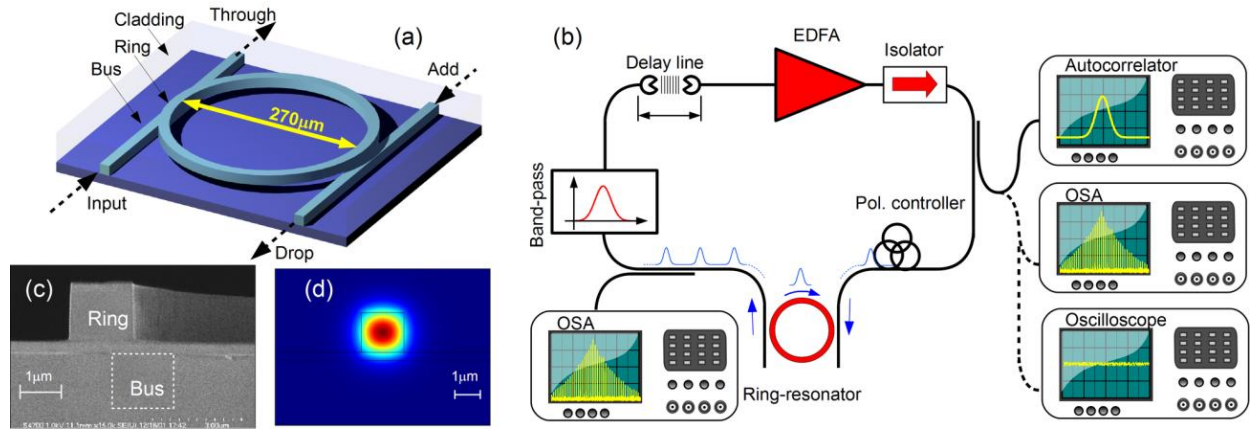

**Figure 1. Filter Driven Four Wave Mixing design.** **(a)** Schematic of the central component – a monolithically integrated 4-port high-Q (Q=1.2 million) micro-ring resonator (fibre pigtails not shown) **(b)** High repetition rate laser based on Filter-Driven Four-Wave-Mixing. **(c)** SEM picture of the ring cross section before depositing the upper cladding of $SiO_2$. The waveguide core is made of high index (n=1.7) doped silica glass. **(d)** Electric field modal distribution for a TM polarized beam calculated through vectorial mode-solving.

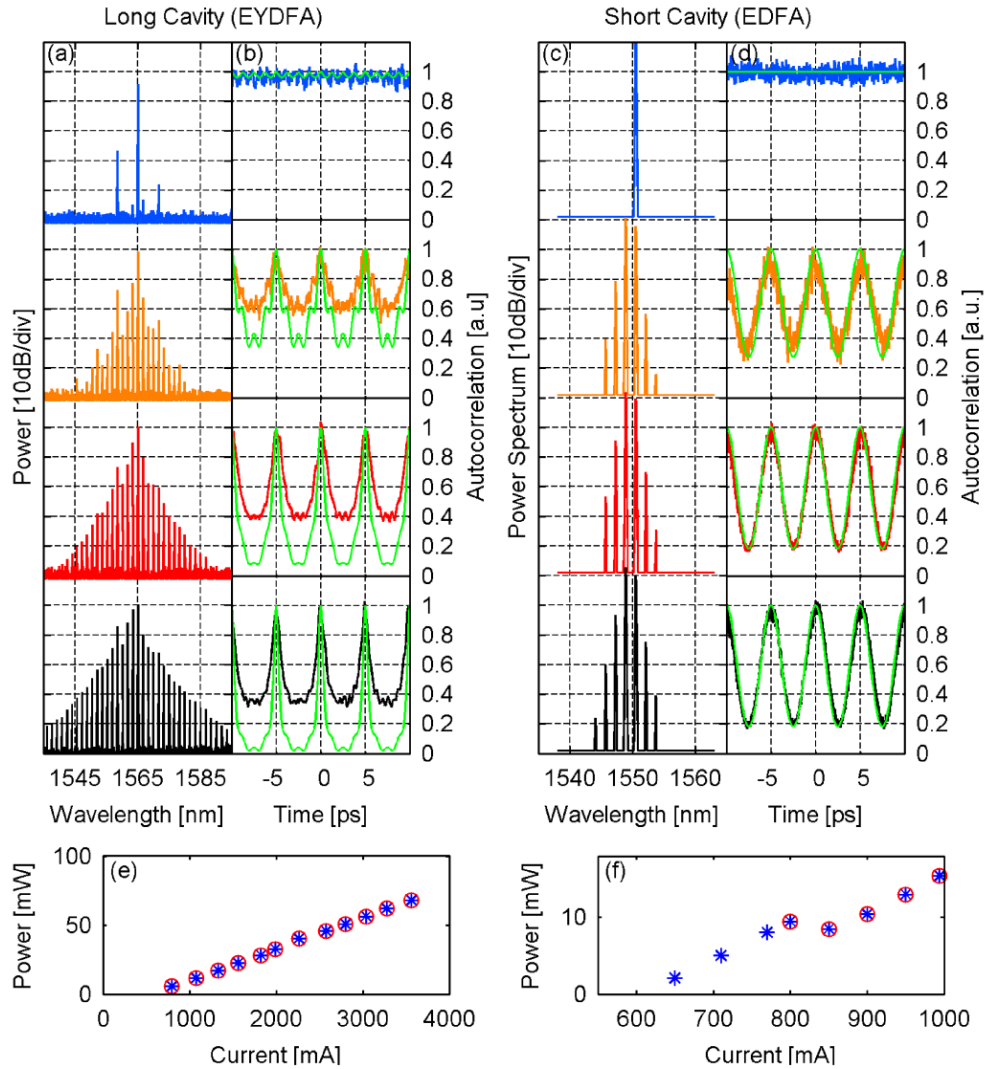

**Figure 2. Experimental optical spectra and autocorrelation traces of the laser output, for different FSRs of the main cavity.** **(a-b)** Long cavity (FSR=6.0MHz) laser emission for increasing (top to bottom) pump powers, achieved through the use of a long EYDFA, for 5.5, 28, 40 and 68mW average ring input powers, respectively. In (b) the autocorrelation traces calculated starting from the experimental optical spectra for a fully coherent and transform limited system are also shown in green. These profiles are calculated by considering each line of the experimental optical spectra as being perfectly monochromatic and in-phase with the others, yielding an output pulse with a width (FWHM) of 730fs (duty cycle $\rho$=0.15) for the highest excitation power condition. The measured autocorrelation shows a considerably higher background (the peak-to-background ratio is 2.5:1 for the 68mW case) than the expected autocorrelation (50:1). **(c-d)** Short-cavity laser emission using a short-length, low power EDFA. The average powers in the ring were as low as 8, 10.4, 13.0 and 15.4mW, respectively, with the main cavity FSR=68.5MHz. Due to the lower cavity energy achievable using the short-length EDFA, we show here a maximum spectral bandwidth limited to 7 or 8 lasing modes associated with the 200GHz FSR of the ring resonator. The

autocorrelation traces expected from the optical spectra for a fully coherent transform limited system, shown in green, perfectly matches the measured traces, corresponding to a transform-limited pulse width (FWHM) of 2.3ps (duty cycle ρ=0.46), with a peak to background ratio of 5:1 for the 15.4mW case. In all cases the oscillation condition was self-starting and did not critically depend on the initial conditions. (e-f) Average input power in the ring vs. the driving current for the EYDFA and the EDFA, respectively. Red circles indicate multi-wavelength oscillation. In the EDFA case the delay line was adjusted to keep the system in stable operation for driving currents >850mA.

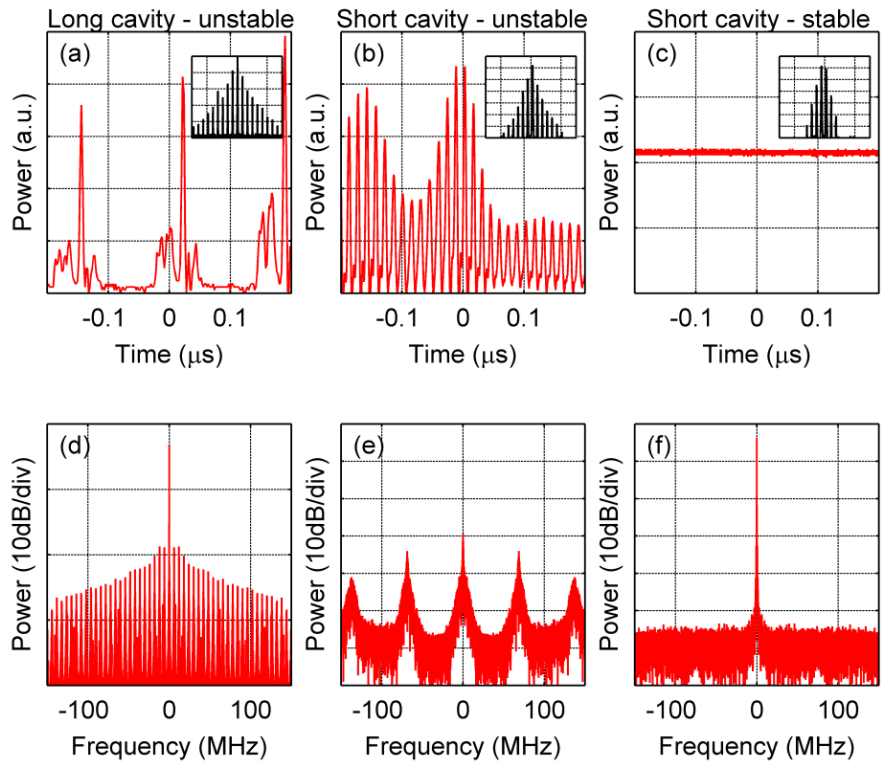

**Figure 3. Laser RF Noise. Top: output intensity. Bottom: RF spectrum. (a) and (d)** Unstable oscillation condition for a main cavity FSR=6MHz, obtained using the EYDFA (average ring input power 68mW) and corresponding to the results shown in Figure 2 (a). The RF signal clearly shows an irregular pulsation. This is reflected in the appearance of harmonics in the RF spectrum at 6MHz intervals. The limited bandwidth of the RF spectrum is consistent with the linewidth (160MHz) of each ring resonator line. **(b) and (e)** Unstable oscillation condition for FSR=68.5MHz obtained using the EDFA (multiple cavity modes oscillating per filter resonance). In this configuration the long timescale output shows complex temporal behaviour with the RF spectrum containing peaks corresponding to 68.5MHz-spaced harmonics. **(c) and (f)** Stable oscillation condition for FSR=67MHz, corresponding to the case of Figure 2 (c) (one cavity mode oscillating per filter resonance). The average power in the ring was 15.4mW. The black inset shows the optical spectrum. This regime is obtained by adjusting the phase of the cavity modes relative to the ring resonator modes via a free space delay line (Figure 1).

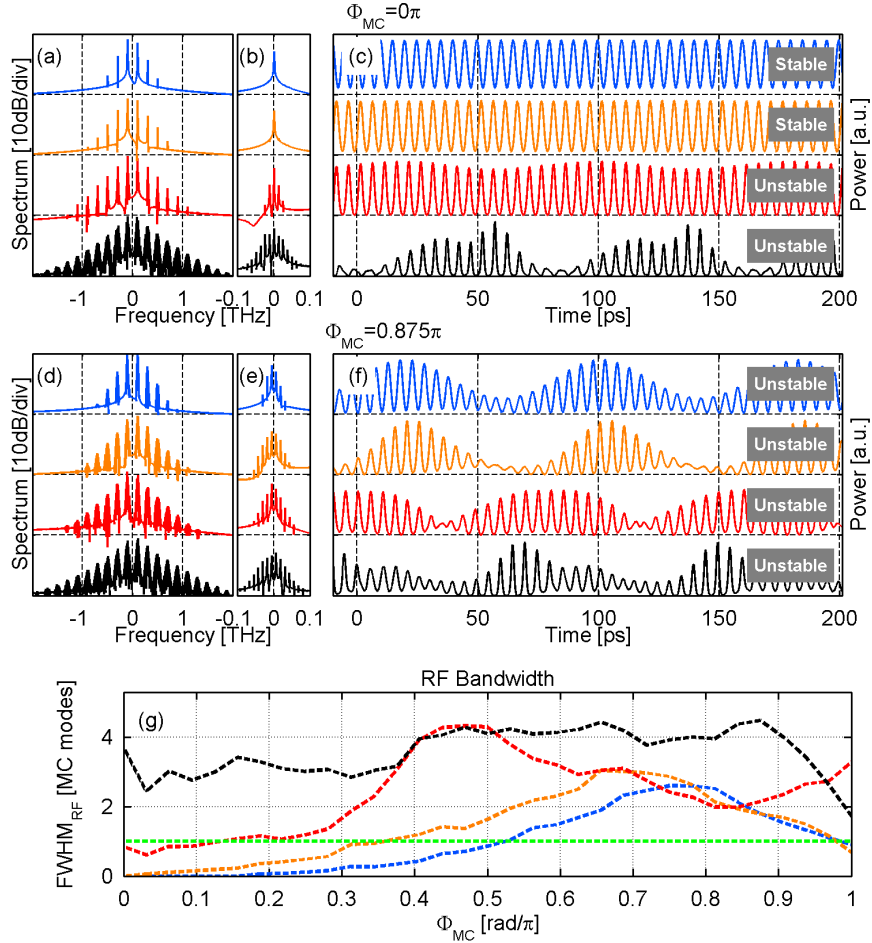

**Figure 4 Theoretical dependence of the laser stability on the amplifier saturation energy and phase $\Phi_{MC}$ of the main cavity modes.** In all the plots $\Phi_{RC} = \pi$ (even number of oscillating ring resonances). The blue, orange, red and black plots are for growing cavity energy, respectively 1,2,3,4 times the amplifier saturation energy associated to the blue line. **(a-c)** Lasing condition for $\Phi_{MC} = 0$. **(a,b)** Optical spectrum **(c)** Temporal evolution, showing modulated peaks in the unstable cases. **(d-f)** Lasing condition for $\Phi_{MC} = 0.875\pi$. **(g)** RF bandwidth (FWHM) in function of the main cavity modes phase $\Phi_{MC}$, expressed in units of number of main cavity modes (i.e., as the ratio between the RF bandwidth and the main cavity FSR). The plots **(b,e)** show a zoom on the central ring lines in **(a,d)**, where the main cavity modes are clearly distinct (corresponding to only one line in the stable case). For larger excitations, the number of oscillating main cavity modes increases.

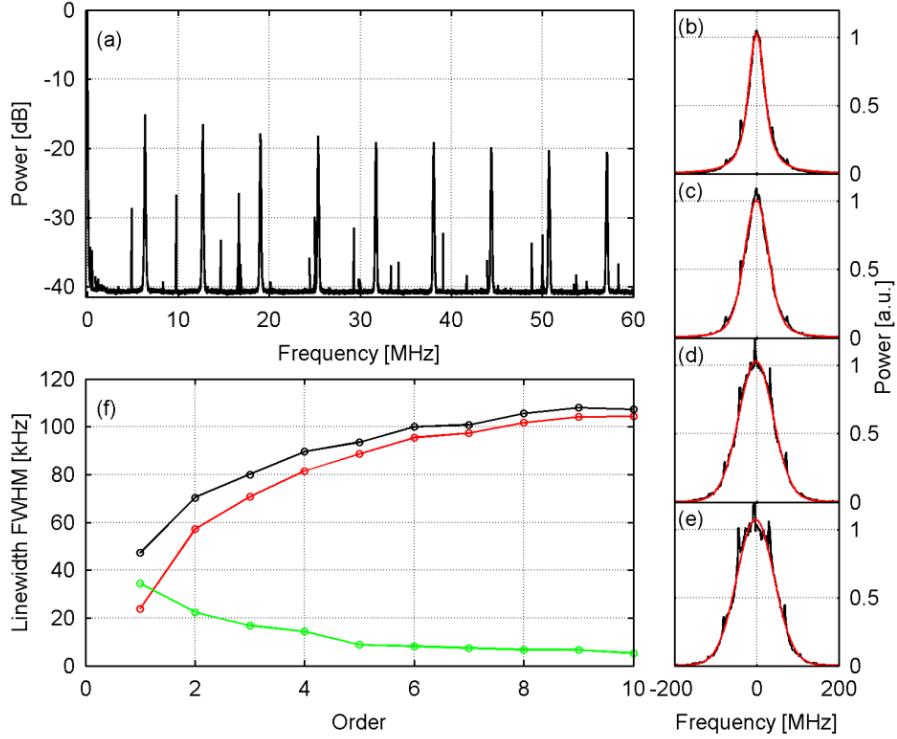

**Figure 5. Phase noise measurement performed by way of a re-circulating delayed self-heterodyne interferometer (RDSHI) with a delay line of 5.8km, and a modulation frequency of 6.5MHz** [7,24,30-32]. The setup measurement is described in Figure 1 or ref. [30]. (a) Broadband view of the RDSHI photocurrent power spectrum. The recirculating beat notes are separated by the modulation frequency of 6.5MHz. (b-e) RDSHI photocurrent power spectrum of the $1^{st}$, $2^{nd}$, $6^{th}$, and $10^{th}$ beat note (black) along with a bestfit to the Voigt function. The Voigt function is defined as the convolution of a Gaussian and a Lorentian function (representing the extrinsic and intrinsic contribution to the noise respectively), e.g. ref [24] equation 1. (f) Linewidth FWHM (black dots), Lorentian (green dots) and Gaussian (red dots) contributions of the Voigt best fit function versus the order of the beat note. From this measurement it is clear that the Lorentian contribution (i.e. the quantum limit for the linewidth) saturates to a value well below 13 kHz, while the Gaussian contribution is dominant for the largest order, indicating that the noise is dominated by an extrinsic/environmental contributions.